\documentclass[9pt,twocolumn,twoside]{ilcss}

\templatetype{ilcssworkingpaper} 

\title{Patterns in Conflict Dynamics in  Yemen and Syria}

\author{\href{https://github.com/samarmoussa53-web}{Samar Moussa Abdou}}
\author{\href{https://physics.columbian.gwu.edu/neil-johnson}{Neil F. Johnson}}
\affil{\href{https://donlab.columbian.gwu.edu/}{Dynamic Online Networks Laboratory}, George Washington University, Washington, DC 20052, U.S.A }

\leadauthor{Moussa and Johnson} 

\authorcontributions{S.M.A. designed the study, performed the data analysis, and drafted the manuscript. N.F.J. developed the theoretical framework, contributed to interpretation of results, and supervised the research. Both authors discussed the results, edited the manuscript, and approved the final version. }
\equalauthors{\textsuperscript{1}Samar Moussa Abdou and Neil Johnson contributed equally to this work}
\correspondingauthor{\textsuperscript{2}To whom correspondence should be addressed. neiljohnson@email.gwu.edu}


\begin{abstract}
Abstract 

Conflict fatalities tend to follow heavy-tailed statistical distributions.
A 2005 fusion–fission theory predicts mathematically that
for armed groups operating in dynamically evolving clusters within
a given conflict, the number of fatalities per conflict event will follow an approximate power-law distribution with exponent $\alpha$ 
near $2.5$, with the specific $\alpha$ value offering insight into the
relative robustness of larger versus smaller clusters of fighters in that armed group. Since Yemen and Syria are current hotspots for future conflict, yet their most recent conflicts (2023–2025) have not been studied at the event level, 
we use ACLED data 
to determine their best-fit $\alpha$ value as each conflict evolved. We find that $2.5<\alpha<3.5$ predominantly throughout each conflict, which suggests
that the fighters in each of these conflicts continued to operate in smaller clusters as the conflict evolved. Moreover, temporary reductions in the $\alpha$ value -- which suggests a temporary increase in the robustness and involvement of larger clusters of fighters -- appear to arise during major crises ahead of the largest battles. Though the lack higher-quality data for these conflicts prevents us from  establishing this more firmly, such a temporary reduction in $\alpha$ hints at its potential use as an early-warning signature.

\end{abstract}

\doi{\url{http://ilcss.umd.edu/}}

\usepackage{fancyhdr}

\fancypagestyle{firststyle}{
  \fancyhf{}
  \fancyfoot[C]{\thepage}

}

\AtBeginDocument{\pagestyle{fancy}}

\usepackage{fancyhdr}
\AtBeginDocument{\pagestyle{fancy}}

\begin{document}

\maketitle
\thispagestyle{firststyle}
\ifthenelse{\boolean{shortarticle}}{\ifthenelse{\boolean{singlecolumn}}{\abscontentformatted}{\abscontent}}{}


\section{Introduction}

\dropcap{T}he study of human conflicts and terrorism (no matter how defined) has a long and distinguished history across a wide variety of disciplinary perspectives
\cite{cederman2003modeling, sundberg2013ucdp, gtd2016, goldstone2010forecasting, cederman2017predicting, eck2012comparison, weidmann2013resolution, weidmann2015accuracy, 
silver2012signal, braumoeller2019only,wikimedia_damascus_archway_2025, wikimedia_israeli_attacks_yemen_2024, wikimedia_assad_equipment_2024, acled}. Irrespective of their myriad causes, it is clear that conflicts tend to exhibit highly irregular fatality patterns \cite{richardson1948variation, richardson1960statistics,  johnson2009, johnson2011patterns, johnson2006universal, economist2005rules, johnson2013benchmark, clauset2009powerlaw, johnson2016isis, spagat2018fundamental, spagat2020unifying, johnson2020computational, tkacova2023actors, manrique2023shockwave, huo2025epl, gonzalezval2015warsize, clauset2007terrorism, clauset2013estimating, clauset2010reply, gillespie2014powerlaw, gleditsch2020richardson, clauset2018severity}. 
Instead of following thin-tailed statistical models such as a typical bell-curve 
Gaussian, conflict fatalities routinely form heavy-tailed distributions in which 
rare but extremely deadly events occur far more often than everyday assumptions about the randomness of war 
would predict~\cite{spagat2018fundamental, spagat2020unifying, johnson2009,johnson2020computational,huo2025epl,clauset2009powerlaw}. 
This empirical regularity of a heavy-tailed distribution suggests that severe casualty events may arise from underlying 
organizational mechanisms rather than random fluctuations.

\begin{figure}[tbhp]
\centering
\includegraphics[width=1\linewidth]{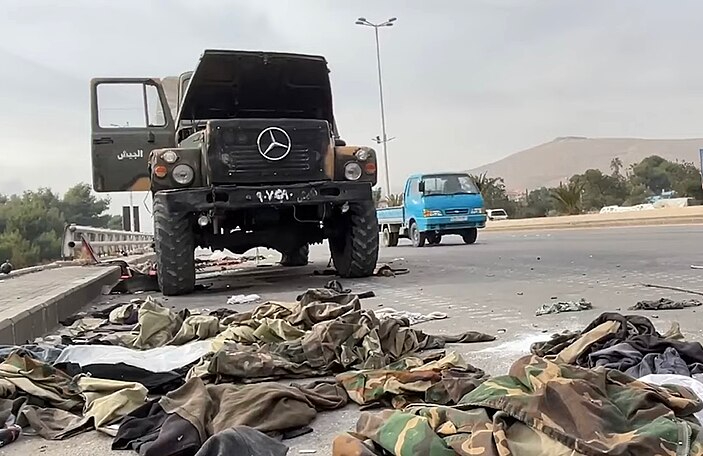}
\caption{A military truck and uniform belonging to former Assad regime soldiers, photographed after the fall of the regime. The abandoned equipment supports the notion that there are moments of rapid fragmentation and collapse of centralized military organization as a conflict evolves. Source: Wikimedia Commons \cite{wikimedia_assad_equipment_2024}.
}
\label{fig:net}
\end{figure}

A recent study \cite{huo2025epl} provides a physics-based theory that can explain quantitatively and mechanistically such fatality patterns in Israel--Palestine regional conflicts, including the most recent war in Gaza that started in October 2023. 
It employs the original mathematical fusion--fission model for conflict and terrorism first published in 2005 on the public arXiv server (\cite{johnson2006universal, economist2005rules,johnson2009,spagat2018fundamental}). 
Its prediction of $\alpha$ being in the vicinity of $2.5$ for any given conflict,
has since been validated in tests on the conflicts that started after the prediction appeared publicly in 2005 -- and on the datasets that were produced after this same 2005 date -- 
which means that each test was strictly out-of-sample  \cite{johnson2009, johnson2006universal, economist2005rules, johnson2013benchmark, johnson2016isis, spagat2018fundamental, spagat2020unifying, johnson2020computational, tkacova2023actors, manrique2023shockwave, huo2025epl}. 
It was also recently shown to explain the well-known cross-war casualty distributions that date back to Richardson (\cite{spagat2020unifying, johnson2009,johnson2020computational,richardson1948variation, richardson1960statistics}) as well as casualties by region within a given country \cite{tkacova2023actors} -- and also across online spaces that nurture the extremist support from which real-world fighters can then emerge  \cite{manrique2023shockwave}. 

This fusion-fission model is admittedly highly unconventional from a social science perspective because it purposely averages over the myriad possible sociological, political, technical, geographical and economic details that might be involved in a given conflict -- yet the model's simplicity is also its strength, as is well-known from similarly minimal models across the physical and biological sciences. By focusing solely on the kinetics of how humans fight collectively, the model yields fusion-fission dynamical equations that can be solved with pen-and-paper using standard algebra, to yield rigorous mathematical predictions regarding the distribution of fatalities (casualties) per event and its expected power-law exponent value $\alpha$ (\cite{johnson2006universal}). 

The independent empirical study shown in Fig. 1 in Ref. \cite{huo2025epl} demonstrates that the fusion-fission fighter mechanism that forms the core of the theory (\cite{johnson2006universal, economist2005rules}), is indeed realistic. 
Specifically, this empirical study shows that fighter groups undergo a continual dynamical organizational 
process during a conflict, in which they fuse into larger coordinated clusters but then occasionally 
fragment into smaller clusters when under pressure, e.g. they may scatter when sensing 
danger. It can then be shown mathematically (\cite{johnson2006universal}) that when fusion dominates fission, the system reaches a stationary regime in which the 
distribution of fighters' cluster sizes follows a power law with exponent $\alpha=2.5+\delta$ derived mathematically from first principles, where $-1\leq\delta\leq 1$ according to the relative robustness of larger clusters of fighters. The amount that the best-fit empirical $\alpha$ value for a given conflict deviates from $2.5$, hence offers insight into the relative stability of larger clusters of fighters in that conflict~\cite{johnson2006universal,johnson2009}.
Specifically, the mathematics of this fusion--fission theory predicts that in a conflict in which 
larger clusters of fighters are relatively more robust (i.e.\ more akin to a formal army’s 
formation on a traditional battlefield), then the empirical $\alpha$ should be closer to 1.5 (i.e. $\delta\rightarrow -1$). 
This is because the presence of such large robust clusters that can do proportional amounts of damage, will tend to mean 
that there will be events with large numbers of fatalities on average: hence there will be a flatter 
tail (i.e.\ smaller $\alpha$). In the opposite limit in which the conflict features 
many smaller fighter clusters and hence smaller numbers of fatalities per event (i.e.\ 
more akin to a guerrilla war or insurgency) then the empirical $\alpha$ should be closer to 3.5 (i.e. $\delta\rightarrow 1$). 
This steeper tail (i.e.\ larger $\alpha$) means that individual conflict events with 
large numbers of fatalities are predicted to be relatively rare. When these two extremes are 
balanced, the prediction is that the empirical $\alpha$ should be approximately equal to $2.5$.

This paper contributes to the current conflict literature in two ways. 
First, we analyze the empirical distributions of fatalities for Yemen and Syria. Neither Yemen nor Syria have been featured in previous analyses of event-level data to our knowledge -- and most importantly, each is a potential conflict hotspot going forward and hence of great interest to the U.S. and its allies. We use the ACLED ~\cite{acled} fatality data for Yemen and Syria, together with the established 
maximum-likelihood power-law fitting~\cite{clauset2009powerlaw} as well as rolling estimates 
of $\alpha$. Second, we use the fusion-fission theory to interpret the temporal variation of the 
best-fit empirical $\alpha$ values as each of these conflicts evolves, and hence implied shifts 
in the fusion--fission dynamics of the armed group.
Such temporal variation of the best-fit empirical exponent $\alpha$ for periods within a given conflict as that 
conflict evolves, has been less studied either empirically or theoretically. 
We freely acknowledge that our exploration of within-conflict variations in $\alpha$ is still strictly speaking tentative because of the need for higher accuracy event data which is not currently available -- and indeed may never become available. However such data issues are a common challenge facing any conflict study, not just Yemen and Syria -- hence like other researchers, we must just move forward with the challenge using the best available data.
\vskip0.1in 

We stress that we do {\em not} focus here on the more subtle question of whether other types of distribution may have equally good, or better fits than the power law -- nor do we focus on scrutinizing specific measures of how well the power-law fits all the data points given the low quality of some of the data reporting. Indeed, it is well known that artificial rounding of reporting in casualties (e.g. reporting 20 versus the true number 18) can quickly degrade the goodness-of-fit statistics -- but it has less effect on the best-fit estimate for the exponent $\alpha$. Taking this into account together with the imperfect nature of current event-level data for Yemen and Syria, we will simply focus this paper on trends in the best-fit empirical exponent $\alpha$ values.
\vskip0.1in 

Qualitatively, it is known that both the Yemen and Syria conflicts began in a seemingly fragmented way where 
violence was dispersed, localized, and dominated by small fatality events. But both conflicts then 
experienced political and military shocks, including major airstrikes, massacres, and 
regime changes that could in principle have reorganized fighters temporarily into larger coordinated formations, before then possibly further fragmenting again (e.g. Figs. 1,2). We might 
therefore expect that such temporary reorganizations should manifest themselves statistically 
as temporary decreases in the exponent value $\alpha$ from an initially high value toward a lower value in some complex, non-monotonic way -- because a  reduction in $\alpha$ corresponds to an increase in the relative robustness and hence involvement of larger clusters of fighters. This is indeed what we will find in our results. 
In both conflicts, we find that the best-fit empirical exponent begins near 3.0; it then varies in a non-monotonic way but at some stage shows a marked reduction in value  toward $2.5$. Our findings  also hint at downward 
drifts occurring prior to periods associated with conflict escalation,  which in turn hints at fusion-driven reorganization and hence larger clusters of fighters becoming more robust -- hence leading to events with higher casualties and hence a sense of escalation.

\begin{figure}[tbhp]
\centering
\includegraphics[width=1\linewidth]{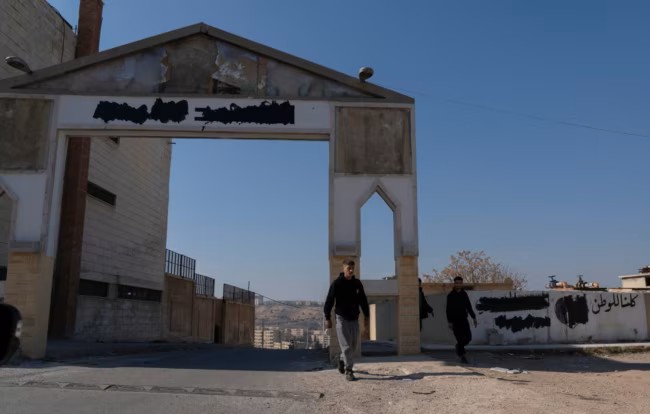}
\caption{An archway in Damascus that previously displayed the slogan “Assad Forever,” shown blacked out following regime collapse in 2025. The image illustrates the symbolic erasure of state authority and the reconfiguration of political space during prolonged conflict. Source: Wikimedia Commons \cite{wikimedia_damascus_archway_2025}}
\label{fig:net}
\end{figure}

\vskip0.1in
Our results are summarized in Figs. 3,4 and Tables 1,2. As well as providing a fresh interpretation of each conflict's evolution, they suggest that  the conflicts in Yemen and Syria -- despite their many social, political etc. differences -- display a crudely common  lifecycle in terms of their casualty statistics and hence $\alpha$ value: an initial state characterized by fragmented fighter clusters, then at some later stage a downward drift corresponding to the formation of robust larger fighter clusters and hence an  escalation of more severe attacks, and then post-crisis 
adjustment. 
More generally, our analysis suggests that the best-fit empirical exponent $\alpha$ could serve as a quantitative tracer of 
organizational dynamics in future conflicts, and can hence provide a principled way to identify when a 
conflict is transitioning into a phase where high severity events (i.e. many casualties) become 
structurally embedded.

\section{ Methodology}

This study analyzes fatality data from ACLED for Yemen (January 2023--September 2025) 
and Syria (January 2023--August 2025). 
Only events with at least one recorded fatality were included. The statistical behavior of interest concerns the distribution of event 
\emph{sizes}, not their occurrence frequency. 
To capture structural changes within the conflict systems, 
each country’s data was segmented into distinct periods reflecting major 
military or political transitions according to broad consensus \cite{acled}. 
For Yemen, these segments correspond to the Saudi border bombardment, 
the Al-Bayda explosion, the Ras~Isa terminal strikes, and the July 2025 battles. 
The Syrian segmentation reflects a prolonged stability period, 
the escalation beginning in late 2024, 
the collapse transition through mid-2025, 
and the post-collapse environment. We also separately analyze a simple 180-day rolling average which is independent of such manual assertions about significant periods within each conflict.

The theoretical motivation for this analysis comes from the 
fusion--fission model of armed-group organization \cite{huo2025epl,johnson2006universal,johnson2009}. 
In this model, clusters of fighters continually fuse and fragment (fusion-fission) as they build up and subsequently scatter when sensing danger. Describing this dynamical fusion-fission process in mathematical form generates the prediction that $\alpha$ will have a value of $2.5+ \delta$ (where $-1\leq \delta\leq 1$ with its precise value  depending on the relative robustness of larger versus smaller fighter clusters in a given conflict). This result can be derived using straightforward college algebra from the following equations for the rate of change of the number of fighter clusters $n_s$ that contain $s$ fighters (i.e. size $s=1,2 \dots N$): 
\begin{align} \label{eq:ez_prod_nat}
 \dot{n}_s 
 &=
\bigg[\substack{
\text{number of fighter clusters of} \\
\text{size $s$ being formed} \\
\text{at time $t$ through}
\\
\text{fusion (or fission)}
}\bigg]
\ -\ 
\bigg[\substack{
\text{number of fighter clusters of} \\
\text{size $s$ removed}
\\
\text{at time $t$ through}
\\
\text{fission (or fusion)}
}\bigg]
& \nonumber \\ 
&=
\bigg[\frac{F}{2}\sum_{\substack{0\leqslant i\leqslant s \\ 0\leqslant j\leqslant s}}^{i+j=s} (ij)^{1-\delta} \frac{n_in_j}{N^2} + \delta_{1s}\frac{\varphi_f}{2} \sum^{\infty}_{j=1}j^{2-\delta} \frac{n_j}{N}\bigg] \nonumber \\
    &- \bigg[\left( F \sum_{i=1}^\infty i^{1-\delta}\frac{n_i}{N} + \frac{\varphi_f}{2} \right) s^{1-\delta}\frac{n_s}{N}\bigg]
\end{align}
where $\varphi_f$ describes the fighter cluster fission probability while $F$ describes the fighter cluster fusion probability. 
$N$ is the number of fighters involved in this fusion-fission and for simplicity we have suppressed additional species labels denoting fighters'  organizations, ideologies or factions (i.e. their type or `species' label). 

We will not repeat again the already-published derivation of $\alpha=2.5+ \delta$ from Eq. 1, or explain it in detail, but instead we refer interested readers to Refs. \cite{huo2025epl} and \cite{johnson2009} where this calculation with all its algebra is laid out in full detail.
We will however provide an interpretation of Eq. 1 in  order to convey its meaning. 

The first parenthesis in Eq. 1 includes all the ways that the number of fighter clusters of size $s$ can increase at a given time $t$: i.e. through fusion of smaller fighter clusters, or for the case of fighter clusters of size $s=1$ (i.e. one lone fighter) the fission of any larger cluster. 
The second parenthesis in Eq. 1 includes all the ways that the number of fighter clusters of size $s$ can decrease at a given time $t$: i.e. through fission of a fighter cluster of size $s$, or their fusion into any larger fighter cluster. 
Specifically, a fighter cluster of size $s=i+j$ is created by fusion of two smaller fighter clusters $i$ and $j$ (e.g. $6+5=11$). A fighter cluster of size $s$ is lost either by its fusion with another fighter cluster or its fission (e.g.  $11=1+1+\dots+1+1$). 
This result $\alpha=2.5+ \delta$ is remarkably robust to variations in the model setup and assumptions. Also its mathematical validity only requires that the post-fission cluster fragments are  mathematically small ( $<s_{\rm min}$), i.e.  it does not actually have to be total fission as in Eq. 1. 
Any temporal variation of $N$ just needs to be slow compared to these fighter cluster dynamics. 
If the fighter force comprises only one type of fighter, $F$ is a number that depends on those fighters' average heterogeneity \cite{huo2025epl}; but with $D>1$ adversarial species (where `species' means a different type, e.g. different organization, or ideology, or allegience, or faction) $F$ becomes a $D$-dimensional matrix. We refer to  \cite{huo2025epl} for full mathematical details and equations. 
\vskip0.1in

Akin to the known science of reaction kinetics, it is reasonable to take the number of fighters $s$ in a cluster as determining -- on average -- the total number of casualties $x$ in an event that involves that fighter cluster, on average. Here $x$ can include fighters on either side and civilians, and it can be any constant multiple or fraction of $s$. 
Hence taking the rate at which fighter  clusters are involved in events as a constant (e.g.  because every fighter needs a similar time to recover or re-equip, regardless of its fighter cluster size) the distribution $n_s$ will on average have the same mathematical form as the casualty distribution $n_x$, i.e.  the number of events with $x$ casualties. Hence the power law $n_s \sim s^{-\alpha}$ for the distribution of fighter cluster sizes also describes the distribution of casualties per event (i.e. event severity) on average. This approximation, though reasonable on a crude level, obviously averages over the fact that in real conflict data, the severity of a given event will be affected by specific weapon types (airstrike vs ground clash), target types (civilians vs combatants), reporting biases, and event aggregation artifacts (multiple incidents coded as one event).
Taken together, this means that -- based on the 2005 fusion-fission model -- the distribution of fatalities per event during a conflict should have an approximate  
power-law distribution
$
p(x)
=
\frac{\alpha - 1}{x_{\min}}
\left(\frac{x}{x_{\min}}\right)^{-\alpha}
$
for all $x \ge x_{\min}$. 
This mathematical finding based on a simple yet empirically realistic mechanistic fusion-fission fighter clustering mechanism, motivates us to consider the power-law distribution exclusively in our analysis.
We will therefore now use this power-law form to obtain best-fit values for $\alpha$ during the Yemen and Syria conflicts. 

\vskip0.1in
We stress again that our goal in this paper is {\em not} to test whether other types of distributions provide better fits to this Yemen and Syria data. Nor do we focus on how good the power-law fits are: the casualty data is far from being perfectly accurate, but it is the best available -- and we proceed with that always in mind. 
The power-law form means a high probability of events with a low number of casualties: but events with a very large number of casualties are still quite possible. 
Values of  $\alpha$ closer to $1.5$ indicate that such extreme events are more likely and can even be considered as structurally expected, 
while values closer to $3.5$ correspond to a lighter-tailed regime 
where large events are statistically suppressed and hence relatively unexpected.

\vskip0.1in
Because of the difficulties of casualty reporting under conflict conditions, crudely rounded numbers (e.g. 20, 50, 100) tend to be  reported for events instead of the actual integer numbers (e.g. 17, 52, 98). Also, the number of casualties per event is typically much larger than 1 in the data ($x\gg 1$). Taken together, this means that we can use for convenience the continuous power-law likelihood. Prior publications show that this is acceptable when $x\gg 1$; and that while casualty number rounding degrades the goodness-of-fit for a power-law, the best-fit estimate for the exponent $\alpha$ can still be accurate \cite{johnson2009}.
We can somewhat mitigate the impact of noise by focusing on the complementary cumulative distribution 
function (CCDF):
\begin{equation}
P(X \ge x)
=
\left(\frac{x}{x_{\min}}\right)^{-(\alpha - 1)}
\end{equation}
which will appear crudely speaking as a rough straight line on log--log axes. However, we stress that our best-fit exponent values $\alpha$ are obtained using the state-of-the-art power law analysis software tools, not straight-line fits on log-log plots which are known to yield biased estimator values.

\vskip0.1in
Specifically, we estimate $\alpha$ using the Clauset--Shalizi--Newman maximum-likelihood method 
\cite{clauset2009powerlaw}. 
The lower cutoff $x_{\min}$ is selected to minimize 
the Kolmogorov--Smirnov (KS) distance between the empirical data 
and the fitted model. 
Given the $n$ tail  observations $x_i \ge x_{\min}$, 
the log-likelihood for the continuous power law is
\begin{equation}
\ln L(\alpha)
=
n\ln(\alpha - 1)
-
n\ln x_{\min}
-
\alpha
\sum_{i=1}^n
\ln\left(\frac{x_i}{x_{\min}}\right)
\end{equation}
and maximizing this expression yields the best-fitting exponent $\alpha$.  
The closed-form continuous estimator for $\alpha$ is given by 
$1 +
n\left[
\sum_{i=1}^{n}
\ln\left(\frac{x_i}{x_{\min}}\right)
\right]^{-1}$.
Goodness-of-fit can be evaluated using the KS statistic
$D = \max_x |S(x) - P(x)|$
where $S(x)$ is the empirical CDF and $P(x)$ is the fitted model. 

\vskip0.1in 
In addition to evaluating the best-fit $\alpha$ values during separate periods within the two conflicts (Figs. 3,4 and Tables 1,2),  
we also compute a simple rolling estimate of $\alpha$ using overlapping time windows of width 180 days
(see Figs. 3,4). 
Short-term fluctuations in the resulting $\alpha$ estimates may reflect sampling noise, 
but sustained trends can in principle reveal organizational change. 
We are also interested in any marked  negative behavior of the derivative $
\frac{d\alpha}{dt} < 0 $ (i.e. a marked drop in $\alpha$) since this would suggest that the fighter system is temporarily reorganizing toward fused 
operational structures during that time-window, and hence large coordinated attacks will become 
structurally more likely. 

\vskip0.1in
We fully recognize that our study of the evolving best-fit $\alpha$ value, and hence its derivative, is crude because of the lack of good data --  and that it would benefit from full statistical significance analysis across a wide range of other conflicts. But this is outside the scope of the present paper.
In particular, future work will test rolling-window estimates of $\alpha$ to see whether they suffer from being too  highly autocorrelated, noisy when tail sample sizes fluctuate, and sensitive to $x_{\min}$ jumps and to a few large events.
This would mean that, apart from a structural reason, the sign of the numerical derivative might flip for purely statistical reasons. We leave this to future work for a thorough analysis.

\vskip0.1in 

Despite the current shortcomings due to the incomplete and imperfect nature of event data reported within the Yemen and Syria conflicts, which hinders more in-depth statistical analysis, our study of the period-specific best-fit $\alpha$ values, the resulting 
rolling exponent curves, 
and the potential derivative-based early-warning indicator, 
offer a framework for identifying and interpreting 
organizational transitions in Yemen and Syria.

\begin{figure*}
    \centering
    \includegraphics[width=\textwidth]{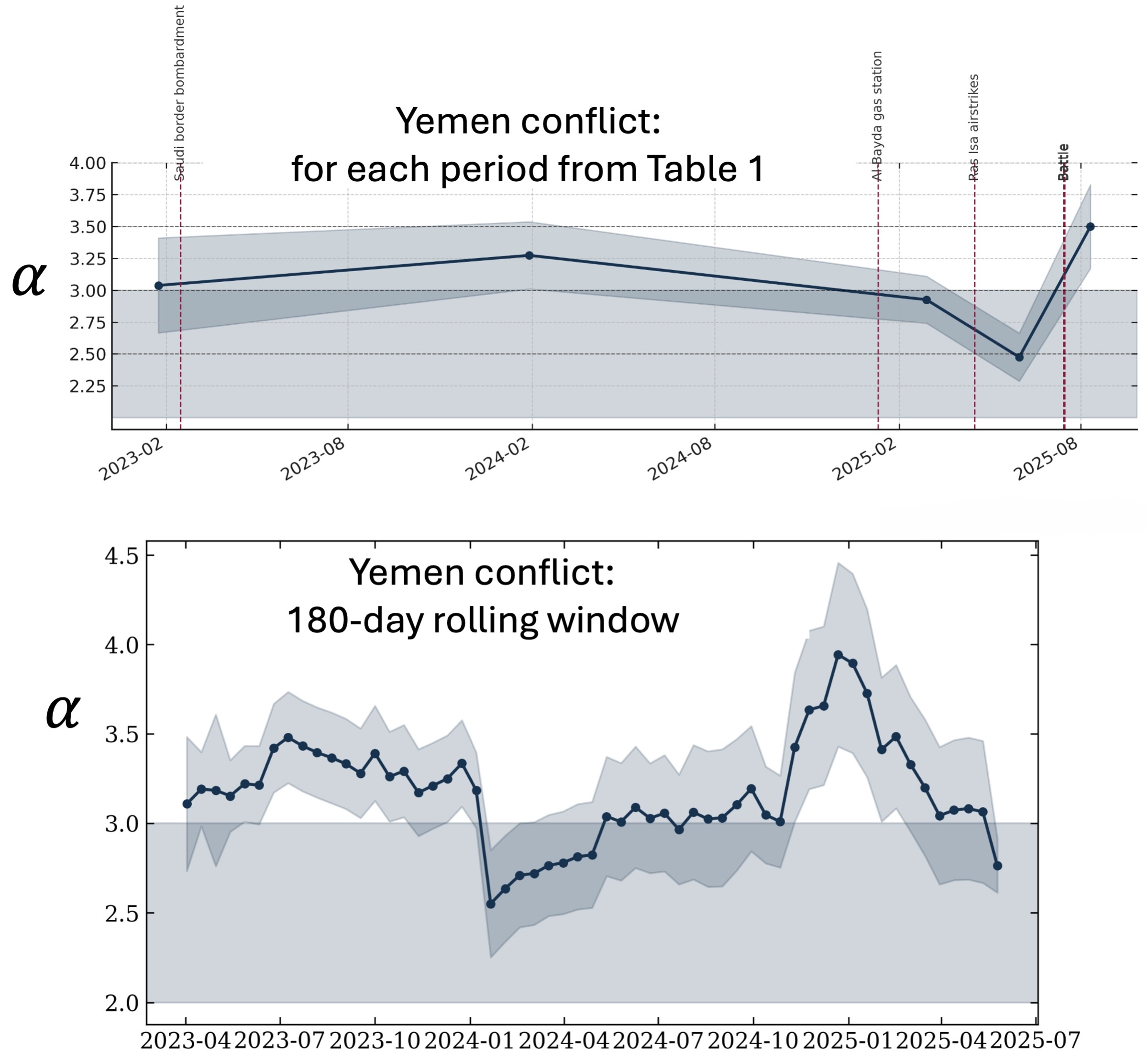}
\caption{
Power-law exponent $\alpha$ estimates for conflict fatalities during the Yemen conflict. Upper panel shows estimates for $\alpha$ within the periods defined by the specific events in Table 1. Vertical lines mark major conflict events. Lower panel shows rolling estimates for $\alpha$ using a 180-day window. Solid lines denote maximum-likelihood estimates with 95\% confidence intervals. As a guide, the shaded band marks $\alpha \in [2.0, 3.0]$. Pronounced downward (or upward) movements in $\alpha$ can be interpreted as periods of heightened (or reduced) coordination and hence events with typically more (or less) casualties. It seems that large downward movements in $\alpha$ can follow major regional shocks -- however this is speculative, and confirming any such association more rigorously will require the availability of higher quality data hopefully in the future. 
}
    \label{fig:placeholder}
\end{figure*}

\section{Results: Yemen (2023--2025)}

Figure 3 and Table 1 suggest that Yemen’s conflict between 2023 and 2025 seems to follow a progression through  
organizational phases. 
During the initial stages, 
violence is fragmented meaning that 
fatality events are small and large coordinated attacks are absent. 
Consistent with this, the estimated $\alpha$ values within these early pre-defined periods as well as from the 180-day moving average are above $3.0$, indicating that events with large numbers of casualties 
are highly unlikely.

A major shock does occur in February 2023 with the Saudi border bombardment. 
However despite this bombardment's scale, the $\alpha$ values do not immediately drop below $3.0$, suggesting that the system absorbs the shock 
without undergoing deeper organizational reconfiguration. The rolling 180-day average shows more variation -- as expected, since it involves a different type of averaging window. It later dips momentarily below $3.0$ toward $2.5$, but then moves back above $3.0$. 
This suggests that overall, the conflict operates in a largely fission-dominated state in which 
fighting units remain dispersed and coordinated large-scale engagements do not 
regularly occur.

In January 2025, the Al-Bayda gas station explosion produces an unusually large 
number of fatalities and is quickly followed by additional high-casualty 
events. 
The exponent for this period decreases toward $\alpha \approx 2.9$, 
and the 180-day rolling estimate of $\alpha$ also shows a persistent downward drift. 
This reduction in $\alpha$ is consistent with the onset of a reorganization phase in which 
conflict actors begin to cluster into larger formations, 
and the system hence transitions to lower $\alpha$ values. 

This drop in $\alpha$ 
suggests that the structural conditions for escalation are forming beneath 
the surface -- and it seems to culminate with the April 2025 U.S. strikes at the 
Ras~Isa terminal and the battles that unfold between April and July 2025. 
During this phase, the period-dependent and the rolling-average
$\alpha$ values both decrease toward $2.5$. 
This lower value of $\alpha$ near $2.5$ suggests that the fighters can now form relatively robust larger clusters and can hence generate events with large numbers of casualties.
High fatality events can in principle arise in rapid succession, reflecting the increased underlying 
cohesion of the fighting units.
The peak of this transformation occurs during the consecutive battles of 
July 15--16, 2025. 
These engagements produce some of the largest fatality counts in the 
entire conflict window. 
At this stage, large coordinated operations drive the dynamics, 
and the conflict appears to exhibit the `ideal' balanced fusion-fission phase in the mathematical fusion--fission model (i.e. $\alpha\approx 2.5$).

\begin{table*}[t]
  \centering
  \caption{Power-law exponent estimates for different periods attached to specific events within the Yemen conflict (2023--2025).}
  \label{tab:yemen_static}
  \begin{tabular}{lclllll}
  \toprule
  Period & Event & $\alpha$ & $x_{\min}$ & Tail $n$ & KS & 95\% CI \\
  \midrule
  2023-01-02 to 2023-02-15 & Saudi border bombardment (~130) & 3.037 & 2.0 & 115 & 0.313 & (2.66, 3.41) \\
  2023-02-15 to 2025-01-11 & Al-Bayda gas station (~40)      & 3.273 & 6.0 & 286 & 0.178 & (3.00, 3.53) \\
  2025-01-11 to 2025-04-17 & Ras Isa airstrikes (~80)         & 2.925 & 2.0 & 421 & 0.308 & (2.74, 3.10) \\
  2025-04-17 to 2025-07-15 & Battle (~100)                    & 2.475 & 2.0 & 236 & 0.296 & (2.28, 2.66) \\
  2025-07-16 to end        & Post–2025-07-16                 & 3.500 & 1.0 & 223 & 0.650 & (3.17, 3.82) \\
  \bottomrule
  \end{tabular}
\end{table*}

After the July 2025 escalation, the period-determined $\alpha$ in the top panel of Fig. 3 seems to suggest that the Yemen conflict transitions back toward higher $\alpha$ and hence more fragmentation -- though the quality of the power-law fit is poor here because of data quality issues, and this result is not reflected in the rolling window $\alpha$ estimates because of this lack of high-quality data.  
If true, this return to a lighter-tailed regime would indicate that the fused 
organizational structure was only temporary and that the system 
re-dispersed following the peak battles.

Overall, the period analysis in Table 1 therefore appears to indicate a crude lifecycle for the Yemen conflict: 
a long initial phase marked by the fragmentation of fighters ($\alpha > 3$); then 
a reorganization period marked by downward drift in $\alpha$; then a high organization phase during mid-2025 
($\alpha \approx 2.5$) featuring several events with large numbers of casualties and hence robust larger fighter clusters;  
and finally a return to a more fragmented phase. 
This structural progression 
aligns with the fusion--fission interpretation. The early phase 
(January--February 2023) shows $\alpha = 3.04$ with a relatively small tail 
and a moderate KS distance, consistent with a fragmented system where 
large-scale coordinated violence is statistically rare. The long period that 
follows (February 2023--January 2025) maintains an even higher 
$\alpha = 3.27$, reinforcing the interpretation that the system absorbs shocks 
(such as the Saudi bombardment) without undergoing a deeper organizational 
reconfiguration.  
The Ras~Isa airstrike period yields $\alpha = 2.93$ with a large tail ($n=421$), 
indicating the first move toward heavy-tailed behavior.  
The most pronounced shift occurs during April--July 2025, where  
$\alpha = 2.48$, corresponding to a robust entry into the universal heavy-tail 
range. In this regime, large fatality events are no longer statistical 
outliers; instead, they reflect more robust large fighting units.
Finally, after July 16, 2025, the exponent rebounds to $\alpha = 3.50$,  
a strong return to fragmentation.  
This rise indicates 
re-dispersion of operational groups and a collapse of fused organizational 
structures. The rolling-window results show more fluctuation -- in part because the 180-day window does not coincide with any specific military period -- but are crudely consistent with this picture.

\begin{figure*}
    \centering
    \includegraphics[width=\textwidth]{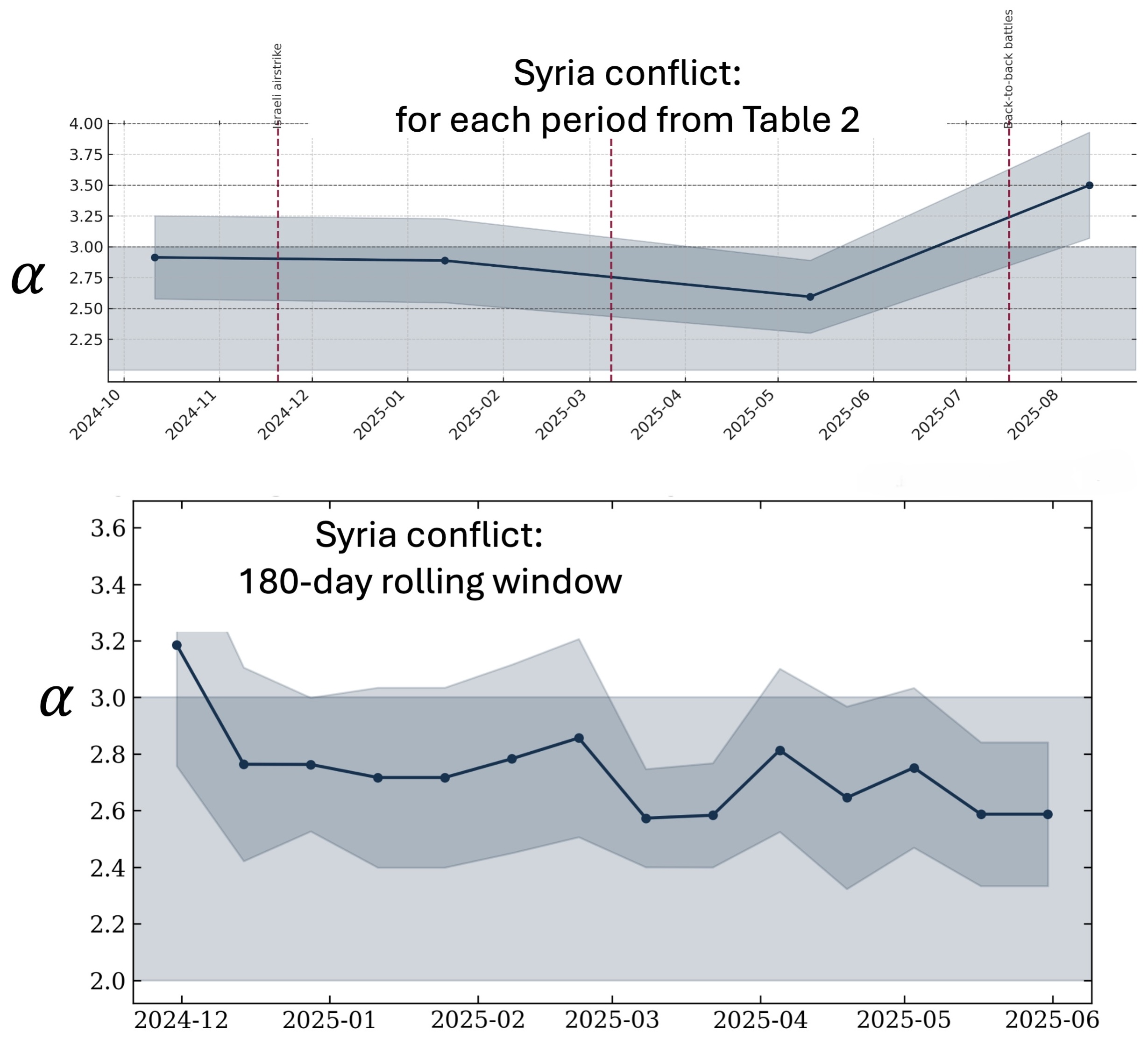}
\caption{
Power-law exponent $\alpha$ estimates for conflict fatalities during the Syria conflict. Upper panel shows estimates for $\alpha$ within the periods defined by the specific events in Table 1. Vertical lines mark major conflict events. Lower panel shows rolling estimates for $\alpha$ using a 180-day window. Solid lines denote maximum-likelihood estimates with 95\% confidence intervals. As a guide, the shaded band marks the regime $\alpha \in [2.0, 3.0]$. }
    \label{fig:placeholder}
\end{figure*}

\section{Results: Syria (2023--2025)}
Figure 4 and Table 2 suggest that Syria’s conflict also follows a progression through  
organizational phases. 
Despite substantial political and social differences from Yemen, 
the $\alpha$ values for the Syrian conflict overall display a crudely comparable lifecycle to those for Yemen: 
an initial fragmented regime with $\alpha$ near $3.0$; then an eventual drop in $\alpha$ to move closer to  $2.5$ prior to the escalation of events with large numbers of casualties; then an apparent increase in the period-determined $\alpha$ to above $3.0$. As for Yemen, data sparcity means that this final rise is not reflected in the final 180-day rolling average value of $\alpha$.

During the initial extended period from January 2023 to November 2024, 
the period-defined $\alpha$ values are near $3.0$, hence events have typically small numbers of casualties and large-scale coordinated 
attacks are uncommon. 
This is 
consistent with reports claiming that  
the regime maintained substantial territorial control, 
while the opposition lacked the cohesion needed to organize large operations. 
Although occasional high-fatality events occurred in the data, most often from airstrikes, 
these remain isolated shocks that do not alter the underlying distribution.
But on November 20 2024, a major Israeli airstrike produced more than one hundred fatalities, 
breaking the statistical pattern of the previous months. 
Unlike similar isolated shocks in Yemen, 
the Syrian system does not re-stabilize after this event. 
Instead, the months that follow include a sequence of unusually large 
incidents, including two regime-linked massacres in March 2025. 
The 180-day rolling estimates of $\alpha$ in time show a decrease (negative derivative) around March 2025. 
This downward trend indicates increasing robustness of larger fighter clusters and hence suggests an early-warning signature of the 
approaching escalation phase.

From March to July 2025, it is known from independent news reports that the Syrian state lost administrative cohesion and 
control over several territories, while opposition groups and local militias 
reconstituted into larger operational units (clusters). 
This structural shift with a subsequent period of back-to-back battles involving robust larger clusters (Table 2) is reflected in both the period-dependent and the 180-day rolling window 
$\alpha$ values sitting nearer to $2.5$. 

\begin{table*}[t]
  \centering
  \caption{Power-law exponent estimates for different periods attached to specific events within the Syria conflict (2024--2025).}
  \label{tab:syria_static}
  \begin{tabular}{lclllll}
  \toprule
  Period & Event & $\alpha$ & $x_{\min}$ & Tail $n$ & KS & 95\% CI \\
  \midrule
  2024-09-01 to 2024-11-20 & Israeli airstrike (~105)            & 2.913 & 4.0  & 124 & 0.209 & (2.57, 3.25) \\
  2024-11-20 to 2025-03-08 & Massacres (~80--94)                 & 2.887 & 11.0 & 118 & 0.092 & (2.54, 3.22) \\
  2025-03-08 to 2025-07-15 & Back-to-back battles (~100)         & 2.594 & 13.0 & 112 & 0.165 & (2.29, 2.88) \\
  2025-07-15 to end        & Post--2025-07-15                    & 3.498 & 5.0  & 130 & 0.277 & (3.06, 3.92) \\
  \bottomrule
  \end{tabular}
\end{table*}

Overall, Syria’s 180-day rolling average $\alpha$ varies less in value than Yemen's, but shows its largest jump at the start: a rapid drop in $\alpha$ indicating the onset of more robust larger clusters in late 2024. 
This is ahead of the 
large-scale reorganizations of early 2025 which eventually lead to a reduction in both the period-determined $\alpha$ and the 180-day rolling average $\alpha$ toward $2.5$.

\section{Conclusion}

The findings of this study provide a fresh perspective on the evolution of recent conflicts in the current political hotspots of Yemen and Syria. More generally, our findings suggest that monitoring the best-fit empirical $\alpha$ over time can provide fresh insight 
into the buildup in robustness of larger fighter clusters which can then lead to periods with more coordinated attacks and hence higher numbers of casualties on average, i.e. conflict escalation. Our findings also help inform the deeper mechanisms shaping conflicts going forward. 

We fully recognize the potential shortcomings of our statistical analysis due to the current lack of highly-reliable casualty event data within the Yemen and Syria conflicts. But with that in mind, we believe that our study of the period-specific best-fit exponent values, 
together with the rolling time-window exponent values 
and the change in these exponents as a potential  early-warning indicator, 
offer a path forward toward identifying and interpreting 
organizational transitions within conflicts in the highly topical hotspots of Yemen, Syria.
Though more work is needed when data becomes available, our results suggest that the exponent $\alpha$ may serve not only as an overall descriptive parameter 
but also as a quantitative tracer of conflict structure and escalation risk in any future conflicts in Yemen and Syria -- and perhaps beyond.

Our findings also show that 
despite major differences in their geography, political context, and military 
composition, the conflicts in Yemen and Syria display a crudely similar statistical 
lifecycle between 2023 and 2025. 
Both conflicts begin in fragmented states characterized by high exponent values 
$\alpha$ near $3.0$, consistent with dispersed fighting units, localized violence, 
and a structural suppression of large fatality events. The $\alpha$ values then at some stage decrease denoting organization into more robust large fighting units. 

Of course, the  $\alpha$ values defined by periods that are attached to specific events (Tables 1 and 2) do not coincide with the $\alpha$ values for the 180-day moving average because they each represent different types of averaging windows in a non-stationary conflict system. But taken together, the Yemen and Syria results suggest that the scaling exponent $\alpha$ 
can provide a unified quantitative lens for understanding the organization of 
violence within modern conflicts. 
However the time-window is defined, the magnitude and the trajectory of $\alpha$ in time can be related not only to an
escalation of larger attacks (i.e. through a decrease in $\alpha$ values and hence more robust larger clusters of fighter and hence more fatalities per event on average) but also the underlying structural shifts driving these 
patterns (e.g. through an increase/decrease in the dominance of fusion over fission which leads theoretically to a decrease/increase in the $\alpha$). 

Though more statistical testing is of course necessary as more accurate data become available, consistent early-warning behavior and the correspondence between 
$\alpha$ and conflict phases suggest that power-law exponents can serve as an 
effective tracer of organizational transitions in armed conflict systems.

\vskip0.5in
\section*{Appendix: Fusion--Fission Generalizations}

This brief discussion does not affect the main empirical results, but helps provide additional theoretical context.
Let $n_s(t)$ denote the number of fighter clusters of size $s$ at time $t$. In the baseline fusion--fission model, fighter clusters evolve through fusion and fission events such that the total number of fighters is approximately conserved: new fighters can appear and others can be removed (killed, injured, retire) as long as the churn is small compared to the overall number of fighters. Fusion events increase fighter cluster size through the merging of smaller fighter clusters, while fission events fragment fighter clusters into smaller units. The mean-field evolution of $n_s(t)$ follows balance equations of the form
\begin{equation}
\frac{dn_s}{dt} = F_s(\{n\}) - D_s(\{n\}),
\end{equation}
where $F_s$ and $D_s$ represent the rates of formation and destruction of fighter clusters of size $s$.
Under broad conditions, the stationary solution yields a power-law fighter cluster-size distribution
$
P(s) \sim s^{-\alpha},
$
with tail exponent $\alpha$ determined by the relative dominance of fusion versus fission processes.
However more generally, fighters may be heterogeneous, belonging to distinct types indexed by $i = 1,\dots,d_S$, representing differences in organization, ideology, or allegiance. In this case, the fighter cluster state is represented by a vector $\mathbf{s} = (s_1, \dots, s_{d_S})$, and fighter cluster populations evolve according to coupled balance equations in a $d_S$-dimensional state space.
The fighter cluster population is then described by a matrix-valued distribution $n_{\mathbf{s}}(t)$, and fusion and fission processes act component-wise on $\mathbf{s}$. Provided post-fission fragments remain small relative to the total system size, the aggregate fighter cluster-size distribution retains a power-law tail, with an effective exponent governed by the same fusion--fission balance as in the single-type case. The net results remains the prediction that $\alpha\sim 2.5+ \delta$, where $-1\leq \delta\leq 1$ with its specific value dependent on the relative robustness of larger fighter clusters within that particular conflict.
Remarkably, many extensions of the model do not alter the primary empirical prediction of a power-law distribution with $\alpha$ in a range around the value $2.5$.

\vskip0.5in

\noindent{\bf Funding} No funding was used in producing this work.
\vskip0.5in

\noindent{\bf Data and Code} All data and power-law testing code are freely available online from the public sources listed in the paper.

\vskip0.5in

\bibliography{ilcss-sample}

\begin{thebibliography}{}

\bibitem[(ACLED), ]{acled}
(ACLED).
\newblock Acled: Armed {C}onflict {L}ocation \& {E}vent {D}ata.
\newblock Available at https://acleddata.com.

\bibitem[Bohorquez et~al., 2009]{johnson2009}
Bohorquez, J.~C., Gourley, S., Dixon, A.~R., Spagat, M., and Johnson, N.~F.
  (2009).
\newblock Common ecology quantifies human insurgency.
\newblock {\em Nature}, 462(7275):911--914.

\bibitem[Braumoeller, 2019]{braumoeller2019only}
Braumoeller, B.~F. (2019).
\newblock {\em Only the Dead: The Persistence of War in the Modern Age}.
\newblock Oxford University Press, USA.

\bibitem[Cederman, 2003]{cederman2003modeling}
Cederman, L.-E. (2003).
\newblock Modeling the size of wars: From billiard balls to sandpiles.
\newblock {\em American Political Science Review}, 97(1):135--150.

\bibitem[Cederman and Weidmann, 2017]{cederman2017predicting}
Cederman, L.-E. and Weidmann, N.~B. (2017).
\newblock Predicting armed conflict: Time to adjust our expectations?
\newblock {\em Science}, 355(6324):474--476.

\bibitem[Clauset, 2018]{clauset2018severity}
Clauset, A. (2018).
\newblock Trends and fluctuations in the severity of interstate wars.
\newblock {\em Science Advances}, 4(2).

\bibitem[Clauset et~al., 2009]{clauset2009powerlaw}
Clauset, A., Shalizi, C.~R., and Newman, M. E.~J. (2009).
\newblock Power-law distributions in empirical data.
\newblock {\em SIAM Review}, 51(4):661--703.

\bibitem[Clauset and Woodard, 2013]{clauset2013estimating}
Clauset, A. and Woodard, R. (2013).
\newblock Estimating the historical and future probabilities of large terrorist
  events.
\newblock {\em Annals of Applied Statistics}, 7(4):1838--1865.

\bibitem[Clauset et~al., 2007]{clauset2007terrorism}
Clauset, A., Young, M., and Gleditsch, K.~S. (2007).
\newblock On the frequency of severe terrorist events.
\newblock {\em Journal of Conflict Resolution}, 51(1):58--87.

\bibitem[Clauset et~al., 2010]{clauset2010reply}
Clauset, A., Young, M., and Gleditsch, K.~S. (2010).
\newblock A novel explanation of the power-law form of the frequency of severe
  terrorist events.
\newblock {\em Peace Economics, Peace Science and Public Policy}, 16(1).

\bibitem[Eck, 2012]{eck2012comparison}
Eck, K. (2012).
\newblock In data we trust? a comparison of ucdp ged and acled conflict events
  datasets.
\newblock {\em Cooperation and Conflict}, 47(1):124--141.

\bibitem[Gillespie, 2014]{gillespie2014powerlaw}
Gillespie, C.~S. (2014).
\newblock Fitting heavy tailed distributions: The powerlaw package.
\newblock {\em Journal of Statistical Software}.

\bibitem[Gleditsch, 2020]{gleditsch2020richardson}
Gleditsch, N.~P., editor (2020).
\newblock {\em Lewis Fry Richardson: His Intellectual Legacy and Influence in
  the Social Sciences}.
\newblock Springer Open.

\bibitem[Goldstone et~al., 2010]{goldstone2010forecasting}
Goldstone, J.~A., Bates, R.~H., Epstein, D.~L., et~al. (2010).
\newblock A global model for forecasting political instability.
\newblock {\em American Journal of Political Science}, 54(1):190--208.

\bibitem[Gonz{\'a}lez-Val, 2015]{gonzalezval2015warsize}
Gonz{\'a}lez-Val, R. (2015).
\newblock War size distribution: Empirical regularities behind conflicts.
\newblock {\em Defence and Peace Economics}.

\bibitem[Huo et~al., 2025]{huo2025epl}
Huo, F.~Y., Johnson~Restrepo, D.~D., Manrique, P.~D., Woo, G., and Johnson,
  N.~F. (2025).
\newblock Physics reveals and explains patterns in conflict casualties.
\newblock {\em EPL}, 151:12001.

\bibitem[Johnson et~al., 2011]{johnson2011patterns}
Johnson, N.~F., Carran, S., Botner, J., Fontaine, K., Laxague, N., Nuetzel, P.,
  Turnley, J., and Tivnan, B. (2011).
\newblock Pattern in escalations in insurgent and terrorist activity.
\newblock {\em Science}, 333(6038):81--84.

\bibitem[Johnson et~al., 2006]{johnson2006universal}
Johnson, N.~F. et~al. (2006).
\newblock Universal patterns underlying ongoing wars and terrorism.
\newblock {\em arXiv preprint arXiv:physics/0605035}.

\bibitem[Johnson et~al., 2013]{johnson2013benchmark}
Johnson, N.~F., Medina, P., Zhao, G., Messinger, D.~S., Horgan, J., Gill, P.,
  Bohorquez, J.~C., Mattson, W., Gangi, D., Qi, H., Manrique, P., Velasquez,
  N., Morgenstern, A., Restrepo, E., Johnson, N., Spagat, M., and Zarama, R.
  (2013).
\newblock Simple mathematical law benchmarks human confrontations.
\newblock {\em Scientific Reports}, 3:3463.

\bibitem[Johnson et~al., 2016]{johnson2016isis}
Johnson, N.~F., Zheng, M., Vorobyeva, Y., Gabriel, A., Qi, H., Velasquez, N.,
  Manrique, P., Johnson, D., Restrepo, E., Song, C., and Wuchty, S. (2016).
\newblock New online ecology of adversarial aggregates: Isis and beyond.
\newblock {\em Science}, 352:1459--1463.

\bibitem[Johnson~Restrepo et~al., 2020]{johnson2020computational}
Johnson~Restrepo, D.~D., Spagat, M., van Weezel, S., Zheng, M., and Johnson,
  N.~F. (2020).
\newblock A computational science approach to understanding human conflict.
\newblock {\em Journal of Computational Science}.

\bibitem[Manrique et~al., 2023]{manrique2023shockwave}
Manrique, P.~D., Huo, F.~Y., El~Oud, S., Zheng, M., Illari, L., and Johnson,
  N.~F. (2023).
\newblock Shockwave-like behavior across social media.
\newblock {\em Physical Review Letters}, 130:237401.

\bibitem[{National Consortium for the Study of Terrorism and Responses to
  Terrorism (START)}, 2016]{gtd2016}
{National Consortium for the Study of Terrorism and Responses to Terrorism
  (START)} (2016).
\newblock Global terrorism database.
\newblock Data file, https://www.start.umd.edu/gtd.

\bibitem[Richardson, 1948]{richardson1948variation}
Richardson, L.~F. (1948).
\newblock Variation of the frequency of fatal quarrels with magnitude.
\newblock {\em Journal of the American Statistical Association},
  43(244):523--546.

\bibitem[Richardson, 1960]{richardson1960statistics}
Richardson, L.~F. (1960).
\newblock {\em Statistics of Deadly Quarrels}.
\newblock Boxwood Press, Pacific Grove, CA.

\bibitem[Silver, 2012]{silver2012signal}
Silver, N. (2012).
\newblock {\em The Signal and the Noise}.
\newblock Penguin.

\bibitem[Spagat et~al., 2018]{spagat2018fundamental}
Spagat, M., Johnson, N.~F., and van Weezel, S. (2018).
\newblock Fundamental patterns and predictions of event size distributions in
  modern wars and terrorist campaigns.
\newblock {\em PLOS ONE}.

\bibitem[Spagat et~al., 2020]{spagat2020unifying}
Spagat, M., van Weezel, S., Johnson~Restrepo, D.~D., Zheng, M., and Johnson,
  N.~F. (2020).
\newblock Unifying casualty distributions within and across conflicts.
\newblock {\em Heliyon}, 6:e04808.

\bibitem[Sundberg and Melander, 2013]{sundberg2013ucdp}
Sundberg, R. and Melander, E. (2013).
\newblock Introducing the ucdp georeferenced event dataset.
\newblock {\em Journal of Peace Research}, 50(4):523--532.

\bibitem[{The Economist}, 2005]{economist2005rules}
{The Economist} (2005).
\newblock Rules of engagement.
\newblock
  https://www.economist.com/science-and-technology/2005/07/21/rules-of-engagement.
\newblock Science and Technology section.

\bibitem[Tkacova et~al., 2023]{tkacova2023actors}
Tkacova, K., Idler, A., Johnson, N.~F., and Lopez, E. (2023).
\newblock Explaining conflict violence in terms of conflict actor dynamics.
\newblock {\em Scientific Reports}, 13:21187.

\bibitem[Weidmann, 2013]{weidmann2013resolution}
Weidmann, N.~B. (2013).
\newblock The higher the better? the limits of analytical resolution in
  conflict event datasets.
\newblock {\em Cooperation and Conflict}, 48(4):567--576.

\bibitem[Weidmann, 2015]{weidmann2015accuracy}
Weidmann, N.~B. (2015).
\newblock On the accuracy of media-based conflict event data.
\newblock {\em Journal of Conflict Resolution}, 59(6):1129--1149.

\bibitem[{Wikimedia Commons contributors},
  2024a]{wikimedia_israeli_attacks_yemen_2024}
{Wikimedia Commons contributors} (2024a).
\newblock Israeli attacks on {Y}emen, {S}eptember 2024.
\newblock Image from Wikimedia Commons (CC BY-SA).

\bibitem[{Wikimedia Commons contributors},
  2024b]{wikimedia_assad_equipment_2024}
{Wikimedia Commons contributors} (2024b).
\newblock Military truck and uniform of former {A}ssad soldiers after the fall
  of the regime.
\newblock Image from Wikimedia Commons (CC BY-SA).

\bibitem[{Wikimedia Commons contributors},
  2025]{wikimedia_damascus_archway_2025}
{Wikimedia Commons contributors} (2025).
\newblock The archway that once read ``{A}ssad {F}orever'' is now blacked out,
  {D}amascus, {S}yria.
\newblock Image from Wikimedia Commons (CC BY-SA).

\end{thebibliography}

\end{document}